
\input harvmac
\sequentialequations

\def\s{{\rm s}} \def\sb{{\rm\bar s}}
\def\d{{\rm d}} \def\D{{\rm D}}
\def\S{{\rm S}} \def\Sb{{\rm\bar S}}
\def\half{\hbox{$1\over2$}}

\Title{DAMTP R92/33}
{\vbox{\centerline{BRS and Anti-BRS Symmetry}
 \vskip4pt\centerline{in Topological Yang--Mills Theory}}}

\centerline{Malcolm J. Perry~ and~ Edward Teo}
\bigskip\centerline{Department of Applied Mathematics and
 Theoretical Physics}
\centerline{University of Cambridge}\centerline{Silver Street}
\centerline{Cambridge CB3 9EW}\centerline{England}

\vskip .8in
\centerline{\bf Abstract}

We incorporate both BRS symmetry and anti-BRS symmetry into the
quantisation of topological Yang--Mills theory. This refines
previous treatments which consider only the BRS symmetry. Our
formalism brings out very clearly the geometrical meaning of
topolo\-gical Yang--Mills theory in terms of connections and
curvatures in an enlarged superspace; and its simple relationship
to the geometry of ordinary Yang--Mills theory. We also discover
a certain SU(3) triality between physical spacetime, and the
two ghost directions of superspace. Finally, we demonstrate how
to recover the usual gauge-fixed topological Yang--Mills action
from our formalism.

\Date{September 92}

\newsec{Motivation}

Gauge theories apparently form the basis of fundamental physics.
Electroweak theory and QCD are examples of Yang--Mills gauge
theories associated with non-Abelian Lie groups. Even general
relativity may be regarded, in a certain sense, as a gauge
theory of the Lorentz group.

The key property of such a gauge theory is that its so-called
gauge fields transform covariantly under transformations
generated by a certain group. When one quantises the classical
gauge theory using the Feynman path integral formalism, one has
to integrate over all gauge fields. However because of this
gauge-invariance, one is summing over redundant degrees of
freedom, thus leading to infinite results. As it turns out,
following the work of Feynman \ref\rFeynman{R.P.~Feynman, Acta
Phys.~Pol. 24 (1963) 697}, DeWitt \ref\rDeWitt{B.~DeWitt,
Phys.~Rev. 162 (1967) 1195; 1239}, Faddeev and Popov \ref\rFP{
L.D.~Faddeev and V.N.~Popov, Phys.~Lett. B25 (1967) 29}, and
others, a way to evaluate properly the path integral is to first
fix the gauge of the action, and then compensate for this
breaking of gauge-invariance by introducing a Jacobian-like
determinant to the measure of the path integral. This term,
popularly known as the Faddeev--Popov determinant, can be written
as a path integral over new anti-commuting fields called ghosts.
It is these unphysical fields which ensure that the resulting
quantum theory is gauge-invariant and unitary.

It was quite by accident when Becchi, Rouet and Stora \ref\rBRS{
C.~Becchi, A.~Rouet and R.~Stora, Phys.~Lett. B52 (1974) 344;
Commun.~Math.~Phys. 42 (1975) 127} discovered a new set of
transformations which leaves the full Faddeev--Popov Lagrangian
invariant. This so-called BRS symmetry contains the original
gauge symmetry, and it has since then been realised that it plays
a fundamental r\^ole in quantum gauge theories, analogous to the
r\^ole of gauge symmetry in the classical theory. This has lead
to our better understanding of the quantisation of gauge theories.
Not only does ``BRS-quantisation'' simplify the heuristic process
of Faddeev--Popov gauge-fixing, it also generalises to situations
where the latter scheme breaks down. Furthermore, it provides a
geometrical picture of the quantum gauge theory, not unlike the
fibre bundle interpretation of classical gauge theories.

A good review of the modern aspects of the BRS symmetry in quantum
gauge theories may be found in ref.~\ref\rBaulieuPR{L.~Baulieu,
Phys.~Rep. 129 (1985) 1}. In fact, our present day understanding
of this symmetry also includes the so-called anti-BRS symmetry,
discovered soon after the BRS symmetry as an additional symmetry
of the Faddeev--Popov Lagrangian. An important result proved in
ref.~\rBaulieuPR\ is that, given any gauge theory whose
infinitesimal transformations build up a closed algebra with
a Jacobi identity, one can always construct the corresponding
BRS and anti-BRS symmetries. These are generated respectively
by the BRS and anti-BRS operators $\s$ and $\sb$, which satisfy
the fundamental nilpotency condition
\eqna\eNilpotent
$$\s^2=\sb^2=0\ ;\eqno\eNilpotent a$$
and anti-commute with each other:
$$\s\sb+\sb\s=0\ .\eqno\eNilpotent b$$
Thus, it is important to realise that in order to completely
characterise any quantum version of a gauge theory, both the
BRS symmetry and the anti-BRS symmetry must be taken into account
on an equal footing.

The particular type of gauge theory that we will be interested in,
in this paper, is topological Yang--Mills theory, whose quantum
theory was first modelled and shown by Witten \ref\rTQFT{
E.~Witten, Commun.~Math.~Phys. 117 (1988) 353}\ to generate the
Donaldson invariants of smooth four-manifolds. The classical
action of this theory is, for any compact gauge group $G$,
\eqn\eClassicalTYM{\int_M\tr\,[F\wedge F]\ ,}
where $F=\d A+A\wedge A$ is the usual Yang--Mills field strength
of the gauge potential one-form $A$. The trace is over the
gauge group indices of $F$, and $M$ is a compact four-manifold.
The action is invariant under arbitrary variations of the gauge
field $A$, and hence describes a topological field theory
\ref\rBBRT{D.~Birmingham, M.~Blau, M.~Rakowski and G.~Thompson,
Phys.~Rep. 209 (1991) 129}.

Baulieu and Singer \ref\rBS{L.~Baulieu and I.M.~Singer,
Nucl.~Phys.~B (Proc.~Suppl.) 5B (1988) 12}, amongst others
\nref\rLP{J.M.F.~Labastida and M.~Pernici, Phys.~Lett. B212
(1988) 56}\nref\rBMS{R.~Brooks, D.~Montano and J.~Sonnenschein,
Phys.~Lett. B214 (1988) 91}\refs{\rLP,\rBMS}, have demonstrated
how to BRS-quantise this classical action, resulting in Witten's
quantum gauge-fixed action. In fact, the BRS-quantisation scheme
is presently the only known way to construct the quantum theory of
\eClassicalTYM, because of its peculiarly large gauge symmetry.
It turns out that three ghost fields (together with their three
associated anti-ghost fields and three Lagrange multiplier fields)
are needed to completely break the symmetry. All the fields
occurring in topological quantum Yang--Mills theory, and their
properties, are listed in Table 1.

%
%
\topinsert
\def\singleline{\smallskip\hrule\smallskip}
\def\doubleline{\smallskip\hrule\vskip1pt\hrule\medskip}
$$\vbox{
\bigskip
\line{\hfill Table 1.\quad The fields of topological Yang--Mills
      theory  \hfill}
\medskip
\doubleline
\tabskip=1em plus2em minus .5em
\halign to \hsize{$\hfil#\hfil$&\hfil#\hfil&\hfil#\hfil&
      $\hfil#\hfil$&\hfil#\hfil\cr
$ Field $& Meaning & Form Degree &$ Ghost Number $& Statistics \cr
\noalign{\singleline}
A & gauge potential & 1 & 0 & odd \cr
F & field strength & 2 & 0 & even \cr
\noalign{\singleline}
c & usual Faddeev--Popov ghost & 0 & +1 & odd \cr
\bar c & anti-ghost of $c$ & 0 & -1 & odd \cr
b & Lagrange multiplier & 0 & 0 & even \cr
\noalign{\singleline}
\psi & topological ghost & 1 & +1 & even \cr
\bar\chi & anti-ghost of $\psi$ & 2 & -1 & odd \cr
B & Lagrange multiplier & 2 & 0 & even \cr
\noalign{\singleline}
\phi &ghost for ghost $\psi$ & 0 & +2 & even \cr
\bar\phi &anti-ghost of $\phi$ & 0 & -2 & even \cr
\bar\eta & Lagrange multiplier & 0 & -1 & odd \cr}
\doubleline\medskip}$$
\endinsert
%
%

Full details of the construction of this BRS symmetry may be found
in ref.~\rBS. Throughout most of this paper, we will adhere to
the use of differential forms to describe the fields. Thus, $A$
is an anti-commuting field in the sense that it is a one-form,
while the two-form $F$ is even. Also recall from ref.~\rBS\
that $\bar\chi$ and $B$ are both self-dual two-forms, by choice
of gauge-fixing.

Looking back at Table 1 again, a few asymmetries should catch the
reader's eye. Firstly, notice that the topological ghost $\psi$
is a one-form, while its anti-ghost $\bar\chi$ is a self-dual
two-form. Naively, one would expect the anti-ghost to have the
same form degree as its corresponding ghost field, just as in the
$c$--$\bar c$ and $\phi$--$\bar\phi$ systems. The other eye-sore is
that the Lagrange multiplier field $\bar\eta$ having ghost number
$-1$, is quite without a counterpart with ghost number $+1$ and
the same form degree. Why should this be the case?

We claim that these asymmetries appear because only the BRS
symmetry, and not the anti-BRS symmetry, has been built into the
topological Yang--Mills theory. This is perhaps not too
surprising a reason, in view of our remarks earlier in the paper,
that the anti-BRS symmetry necessarily coexists with the BRS
symmetry. In this paper, we will introduce both the BRS and
anti-BRS symmetry into topological Yang--Mills theory, and
recover Witten's action just as Baulieu and Singer did using
only the BRS symmetry. This is not just an unnecessary and
pedagogical exercise. Apart from resolving the asymmetries
noted above, it would also bring into full glory, the geometrical
meaning of topological Yang--Mills theory in terms of connections
and curvatures in an enlarged superspace. Furthermore, in this
formalism, the relationship between the topological and the
ordinary Yang--Mills theories would become so simple, it would
seem hard to believe that topological Yang--Mills theory had not
been discovered earlier within the context of BRS and anti-BRS
symmetry of ordinary Yang--Mills theory.

\newsec{Construction of anti-BRS symmetry}

Recall that the BRS symmetry that Baulieu and Singer \rBS\
constructed rests upon the two gauge fields $A$ and $F$; and
the three ghost fields $c$, $\psi$ and $\phi$. The action of the
BRS operator $\s$ on these five fundamental fields is given by
\eqnn\eSTranfs
$$\openup1\jot \tabskip=0pt plus1fil
\halign to\displaywidth{\tabskip=0pt
  $\hfil#$&$\hfil{}#{}$&
  $\hfil#$&${}#\hfil$&
  $\hfil#$&$\hfil{}#{}$\tabskip=0pt plus1fil&
  \llap{#}\tabskip=0pt\cr
\s A &&& =\psi & - & \D c\ , &\cr
\s c & + & \half[c,c] & =\phi\ , &&&\cr
\s\psi & + & [c,\psi] & = & - & \D\phi\ , &\eSTranfs\cr
\s\phi & + & [c,\phi] & =0\ , &&&\cr
\s F & + & [c,F] & = & - & \D\psi\ , &\cr}$$
where [ , ] is understood to mean the graded bracket, and
$\D\equiv\d+[A,\cdot\,]$ is the gauge covariant derivative.
This set is further supplemented by the s-transformations on
their associated anti-ghosts and Lagrange multiplier fields:
\eqn\eSTranfsI{\eqalign{\s\bar c&=b\ ,\cr \s\bar\chi&=B\ ,\cr
   \s\bar\phi&=\bar\eta\ ,}
\qquad\eqalign{\s b &=0\ ,\cr \s B&=0\ ,\cr \s \bar\eta&=0\ .}}
Note that $\s$ raises the ghost number of its operand by $+1$,
and it anti-commutes with $\d$. It is also easy to verify from
these equations that $\s^2=0$.

Let us now postulate the existence of an anti-BRS operator $\sb$
with ghost number $-1$; and in addition to the above five
fundamental fields, three anti-ghost fields $\bar c$, $\bar\psi$
and $\bar\phi$, corresponding to the ghost fields $c$, $\psi$ and
$\phi$ respectively. These anti-ghosts have the same form degree
as their corresponding ghosts, but have ghost numbers that are
opposite in sign. Then by an obvious mirror symmetry to the
$\s$-transformations \eSTranfs, we demand that the following
$\sb$-transformations hold:
\eqnn\eSbTranfs
$$\openup1\jot \tabskip=0pt plus1fil
\halign to\displaywidth{\tabskip=0pt
  $\hfil#$&$\hfil{}#{}$&
  $\hfil#$&${}#\hfil$&
  $\hfil#$&$\hfil{}#{}$\tabskip=0pt plus1fil&
  \llap{#}\tabskip=0pt\cr
\sb A &&& =\bar\psi & - & \D\bar c\ , &\cr
\sb\bar c & + & \half[\bar c,\bar c] & =\bar\phi\ ,
  &&&\cr
\sb\bar\psi & + & [\bar c,\bar \psi] & = & - & \D\bar\phi\ , &
  \eSbTranfs\cr
\sb\bar\phi & + & [\bar c,\bar\phi] & =0\ , &&&\cr
\sb F & + & [\bar c,F] & = & - & \D\bar\psi\ . &\cr}$$

Similar to the $\s$ case, we have that $\sb^2=0$. Furthermore,
we want $\s$ and $\sb$ to anti-commute, that is $\s\sb+\sb\s=0$.
In fact, one can easily verify that
\eqn\eSSbTranfs{\eqalign{
(\s\sb+\sb\s)A &= \s\bar\psi + \sb\psi + [c,\bar\psi]
  + [\bar c,\psi] + \D\left(\s\bar c + \sb c
  + [c,\bar c]\right)\ ,\cr
(\s\sb+\sb\s)F &= \D\left(\s\bar\psi + \sb\psi + [c,\bar\psi]
  + [\bar c,\psi]\right) + \left[F,\s\bar c + \sb c
  + [c,\bar c]\,\right]\ .}}
In order to make both of these expressions vanish, we must have
as the most general possibility, that
\eqn\eLambdaI{\eqalign{\s\bar c + \sb c + [c,\bar c] &= \lambda\ ,
  \cr \s\bar\psi + \sb\psi + [c,\bar\psi] + [\bar c,\psi]
  &= -\D\lambda\ ,}}
where $\lambda$ is an even scalar field which has vanishing ghost
number. The action of $\s$ and $\sb$ on $\lambda$ may be derived
by imposing the condition that $\s\sb+\sb\s$ acting on $c$ and
$\bar c$ vanishes:
\eqn\eLambdaII{\eqalign{
(\s\sb+\sb\s)c &= \s\lambda + [c,\lambda] + \sb\phi +
  [\bar c,\phi] = 0\ ,\cr
(\s\sb+\sb\s)\bar c &= \sb\lambda + [\bar c,\lambda] +
  \s\bar\phi + [c,\bar\phi] = 0\ .}}
Having made these observations, it is merely routine to check
that $\s\sb+\sb\s$ annihilates all the other fields, so that
\eNilpotent{}\ is valid.

To summarise, we have so far identified nine fields, associated
with a closed BRS and anti-BRS symmetry, and whose generators
$\s$ and $\sb$ are nilpotent. It is interesting to plot the
form degree of these nine fields against their ghost numbers,
in which results in a suggestive pattern, as in Fig.~1. For now
we take ${\cal D}$ to generically denote an operator which
increases by $+1$ the form degree, and thus acts in the upward
direction. Analogously, ${\cal S}$ and $\bar{\cal S}$ are
operators which act toward the right and left respectively,
raising and lowering the ghost number by one unit. We will say
a few words about the significance of this pattern later on.

%
%
\topinsert

\def\mapright#1{\smash{
    \mathop{\longrightarrow}\limits^{#1}}}
\def\mapleft#1{\smash{
    \mathop{\longleftarrow}\limits^{#1}}}
\def\mapup#1{\Big\uparrow
  \rlap{$\vcenter{\hbox{$\scriptstyle#1$}}$}}
\def\diagram{$\matrix{\cr &&&& F &&&& \cr\cr
  && \bar\psi && A && \psi && \cr &&&& \mapup{\cal D} &&&&\cr
  \bar\phi && \bar c & \mapleft{\bar{\cal S}} & \lambda &
    \mapright{\cal S} & c && \phi \cr\cr}$}
$$
\vbox{
\bigskip
\offinterlineskip
\halign{&\vrule#&\strut\quad\hfil#\quad\cr
\omit&$\matrix{2\cr\cr1\cr\cr0\cr}$&&\diagram&\omit\cr
\noalign{\hrule}
\omit&\omit&height10pt&\omit\cr
\omit&\omit&
 &$-2$ \hskip16pt $-1$ \hskip28pt $0$\hskip28pt $+1$
 \hskip16pt $+2$&\omit\cr
\omit&\omit&height8pt&\omit\cr
\noalign{\vskip10pt}
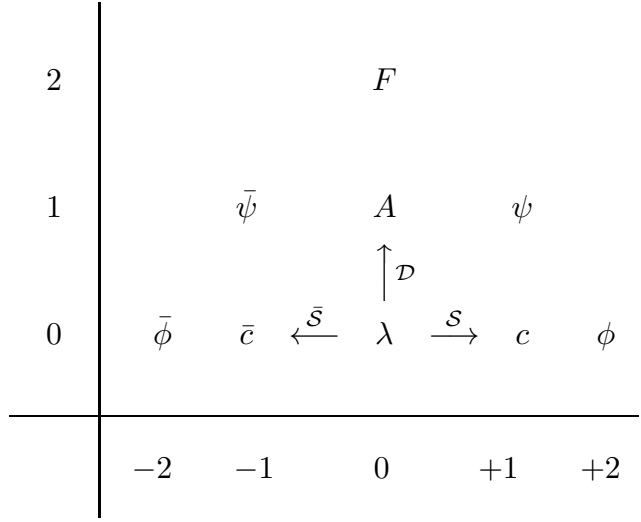
\omit&\multispan5\hfil Fig.~1\quad Plot of form degree
 vs.~ghost number\hfil\cr}
\bigskip}
$$
\endinsert
%
%

\newsec{Geometrical interpretation}

Let us now introduce an even--odd grading of the fields, according
to whether the form degree plus ghost number of the field is even
or odd. So our odd fields are $A$, $c$ and $\bar c$; while the
remaining six fields $F$, $\psi$, $\bar\psi$, $\phi$, $\bar\phi$
and $\lambda$ are even. Note that the operators $\d$, $\s$ and
$\sb$ are all odd. This grading generalises that of ref.~\rBS,
which is simply form degree plus ghost number; sufficient to
classify the fields $A$, $c$, $F$, $\psi$ and $\phi$ only.

Since fields having the same grading are considered to be
essentially of the same nature, we can add them together.
Consider the two expressions
\eqna\eBianchiI
$$\eqalignno{(\d+\s+\sb)(A+c+\bar c) &+
 \half[A+c+\bar c,A+c+\bar c] \cr &\hskip0.93in =
 F+\psi+\bar\psi+\phi+\bar\phi+\lambda\ ,  &\eBianchiI a\cr
(\d+\s+\sb)(F+\psi+\bar\psi&+\phi+\bar\phi+\lambda) \cr
 &+[A+c+\bar c,F+\psi+\bar\psi+\phi+\bar\phi+\lambda] = 0\ .
 &\eBianchiI b}$$
Upon expanding these equations out and collecting terms in form
degree and ghost number, we recover all of the equations
\eSTranfs, \eSbTranfs, \eLambdaI\ and \eLambdaII. That all
these equations may be expressed so compactly in the two
equations of \eBianchiI{}\ is not just a lucky coincidence. But
how can we appreciate the significance of this?

The key \rBaulieuPR\ is to enlarge spacetime $\{x^\mu\}$ into
a superspace ${\cal M}$, with two additional unphysical,
anti-commuting coordinates $\theta$ and $\bar\theta$ at each
point $x^\mu$. Thus, ${\cal M}$ has local coordinates $\{x^\mu,
\theta,\bar\theta\}$, and we can proceed to define differential
forms over this space. The generalised Yang--Mills gauge potential
may be written as the one-form
\eqn\eAi{\tilde{\cal A}(x,\theta,\bar\theta)=\tilde A_\mu(x,
 \theta,\bar\theta)\d x^\mu + \tilde A_\theta(x,\theta,
 \bar\theta)\d\theta + \tilde A_{\bar\theta}(x,\theta,
 \bar\theta)\d\bar\theta\ .}
We will be only interested in fields restricted to the physical
plane, whereby $\theta=\bar\theta=0$. Such a field will be written
without the tilde on the top. Observe that we can make the
identifications
\eqn\eAii{A=A_\mu\d x^\mu\ ,\quad c=A_\theta\d\theta\ ,
 \quad\bar c=A_{\bar\theta}\d\bar\theta\ ,}
so that the generalised Yang--Mills gauge potential over the
physical spacetime is
\eqn\eAiii{{\cal A}=A+c+\bar c\ ,}
precisely the combination of fields occurring in \eBianchiI{}.
This means that the fields $c$ and $\bar c$ can be interpreted
as components of the gauge potential ${\cal A}$ in the unphysical
directions $\theta$ and $\bar\theta$ respectively.

We can also define the analogue of the usual spacetime exterior
derivative by
\eqn\edOperator{\tilde\d=\d+\s+\sb\ ,}
where
\eqn\edOperatorI{\d\equiv\d x^\mu{\del\over\del x^\mu}\ ,\quad
\s\equiv\d\theta{\del\over\del\theta}\ ,\quad
\sb\equiv\d\bar\theta{\del\over\del\bar\theta}\ .}
Thus, in this superspace interpretation, $\s$ and $\sb$ are
exterior derivative operators along the unphysical directions
$\theta$ and $\bar\theta$ respectively.

The curvature two-form or Yang--Mills field strength associated
with ${\cal A}$ is defined in the usual fashion:
\eqn\eFi{{\cal F}\equiv\tilde\d{\cal A}+{\cal A}\wedge{\cal A}=
\tilde\d{\cal A}+\half[{\cal A},{\cal A}]\ .}
{}From \eBianchiI a, we can immediately make the identification
\eqn\eFii{{\cal F}=F+\psi+\bar\psi+\phi+\bar\phi+\lambda\ ,}
whence
\eqn\eFiii{\eqalign{F&=\half F_{\mu\nu}\
 \d x^\mu\wedge\d x^\nu\ ,\cr
\phi&=\half F_{\theta\theta}\
 \d\theta\wedge\d\theta\ ,\cr
\bar\phi&=\half F_{\bar\theta\bar\theta}\
 \d\bar\theta\wedge\d\bar\theta\ ,}
\qquad\eqalign{
\psi&=F_{\mu\theta}\ \d x^\mu\wedge\d\theta\ ,\cr
\bar\psi&=F_{\mu\bar\theta}\ \d x^\mu\wedge\d\bar\theta\ ,\cr
\lambda&=F_{\theta\bar\theta}\ \d\theta\wedge\d\bar\theta\ .}}
While $F$ is the usual Yang--Mills curvature in the physical
spacetime, the other five fields represent the curvature
components along the various unphysical directions. Thus
${\cal F}$ given by \eBianchiI a\ is the total Yang--Mills
field strength in superspace. This is the geometrical
interpretation of the fields occurring in topological Yang--Mills
theory.

Indeed, if we define the super covariant derivatives
\eqn\eCovDer{\eqalign{\D&\equiv\d+[A,\cdot\,]\ ,\cr
\S&\equiv\s+[c,\cdot\,]\ ,\cr \Sb&\equiv\sb+[\bar c,\cdot\,]\ ,}}
it can be readily shown, to our expectation, that
\eqn\eCurv{\eqalign{\D^2X&=[F,X]\ ,\cr \S^2X&=[\phi,X]\ ,\cr
 \Sb^2X&=[\bar\phi,X]\ ,}\qquad
\eqalign{(\S\D+\D\S)X&=[\psi,X]\ ,\cr
 (\Sb\D+\D\Sb)X&=[\bar\psi,X]\ ,\cr
 (\S\Sb+\Sb\S)X&=[\lambda,X]\ ,}}
for any field $X$. This is a pleasing consistency check.

The second equation of \eBianchiI{}\ may be thought of as an
extended Bianchi identity in superspace that ${\cal F}$ satisfies,
that is
\eqn\eBianchiII{\tilde\D{\cal F}\equiv\tilde\d{\cal F}
+[{\cal A},{\cal F}]=0\ .}
Note that the two equations of \eBianchiI{}\ together imply
that $\tilde\d^2=0$, which yields the nilpotency condition
\eNilpotent{}, as well as $\d^2=0$, $\d\s+\s\d=0$, etc.

We can now recover ordinary Yang--Mills theory by imposing the
so-called horizontality condition for ${\cal F}$ \rBaulieuPR:
\eqn\eHorizon{{\cal F}=F\ .}
This is tantamount to requiring that the Yang--Mills field
strength vanish along the unphysical directions, that is, the
identical vanishing of the fields $\psi$, $\bar\psi$, $\phi$,
$\bar\phi$ and $\lambda$. This therefore is the very simple
relationship between the geometries of topological and ordinary
Yang--Mills theories. Indeed, it is just as trivial to proceed in
the other direction. The horizontality condition was discovered
many years ago, when people tried to understand the
BRS-quantisation of ordinary Yang--Mills theory within the context
of superspace. If one had then tried to generalise this to the
case of non-vanishing Yang--Mills field strengths in the unphysical
directions, one would have had at hand topological Yang--Mills
theory. This intriguing historical alternative would have then
resulted in the much earlier discovery of the topological theory.
Incidentally, this argument also clearly shows that the
topological theory, loosely speaking, is the ``most general
type'' of Yang--Mills theory possible, in that there is no more
room in superspace for any other direct generalisation.

To summarise, we have enlarged physical spacetime by adjoining
two new but unphysical directions. In this superspace, we have
introduced a generalised gauge potential ${\cal A}$ and its
field strength ${\cal F}$. These fields each consists of the
classical component and extra ghost components coexisting
together. The idea of having this unified treatment is so that
the original gauge invariance of the classical field can just
be relegated to invariant transformations of its ghost part in
the unphysical directions. Hence, when one integrates the action
over the physical subspace, there is no problem with zero modes
of the classical fields.

For the case of the ordinary Yang--Mills theory, the gauge
potential has a simple gauge invariance of the form $\delta
A=\D\Lambda$. It is made into translations in the unphysical
directions of the form $\s A=\D c$ and $\sb A=\D\bar c$, where
$c$ and $\bar c$ are the ghost components of $A$. This caters
for all the gauge invariance the theory possesses. So our final
unified fields in superspace are ${\cal A}=A+c+\bar c$ and
${\cal F}=F$.

The case of topological Yang--Mills theory is slightly more
complicated. The theory has, in addition to the normal gauge
invariance above, a topological symmetry of the form $\delta A
=\Lambda$. This may be regarded as an invariance of $F$, given
by $\delta F=\D\Lambda$. By the same process done with the $A$
field, we push this gauge invariance of $F$ to its ghost
components. Since $F$ is a two-form, it has five extra ghost
components in all.

\newsec{SU(3) triality}

Our construction of the BRS and anti-BRS symmetry in topological
Yang--Mills theory has also revealed the presence of a hidden
symmetry otherwise absent in ordinary Yang--Mills theory. There
seems to exist a strange type of SU(3) triality between the
physical direction, the $\theta$-direction and the $\bar
\theta$-direction of superspace. This can be seen from Fig.~1.
The gauge potential triplet
$$
\matrix{&A&\cr\bar c&&c}
$$
traces out an isosceles triangle in the ghost number--form
degree plane. This triplet may be taken to be the familiar
weight diagram for the fundamental (1,0)-representation of
SU(3), denoted by {\bf 3}. The gauge field strength sextet
$$
\matrix{&&F&&\cr&\bar\psi&&\psi&\cr\bar\phi&&\lambda&&\phi}
$$
is also an isosceles triangle which may be regarded as the
weight diagram for the (2,0)-representation of SU(3), denoted
by {\bf 6}.

The relationship between these two representations, however,
is rather vague to us. {}From SU(3) representation theory,
we have the relation
\eqn\eSU{{\bf 3}\otimes{\bf 3}={\bf 6}_{\rm s}\oplus
\bar{\bf 3}_{\rm a}\ ,}
that is, the tensor product of two fundamental representations
({\bf 3}) splits into symmetric ({\bf 6}) and anti-symmetric
($\bar{\bf 3}$) parts. This probably describes specifically
the equations
\eqn\eCurvI{\eqalign{\d A+\half[A,A]&=F\ ,\cr
 \s c+\half[c,c]&=\phi\ ,\cr
 \sb\bar c+\half[\bar c,\bar c]&=\bar\phi\ ,}
\qquad\eqalign{\s A+\d c+[A,c]&=\psi\ ,\cr
 \sb A+\d\bar c+[A,\bar c]&=\bar\psi\ ,\cr
 \s\bar c+\sb c+[c,\bar c]&=\lambda\ .}}
If we disregard the exterior derivative terms for the moment, we
may take the tensor product $\otimes$ to be the graded bracket.
Taking brackets of the triplet ${\bf 3}$ then yields the sextet
${\bf 6}$ of fields. The anti-commuting triplet $\bar{\bf 3}$
vanishes because the graded brackets occurring here are
anti-commutators, and thus only single out the symmetric parts.
Perhaps there is a way to incorporate the action of the exterior
derivatives into the definition of the tensor product, but we
will not attempt it here.

Let us just mention another curiosity of Fig.~1. Recall that
there we briefly introduced the operators ${\cal D}$, ${\cal S}$
and $\bar{\cal S}$. Explicitly, we could set
\eqn\eCalDi{{\cal D}\equiv\d+[A,\cdot\,]\ ;}
unless when acting on odd fields $(A,c,\bar c)$, in which case
we take
\eqn\eCalDii{{\cal D}\equiv\d+\half[A,\cdot\,]\ .}
In a similar manner, we  set
\eqn\eCalS{\eqalign{{\cal S}&\equiv\s+[c,\cdot\,]\ ,\cr
\bar{\cal S}&\equiv\sb+[\bar c,\cdot\,]\ ,}
\quad\eqalign{&\hbox{or}\cr&\hbox{or}}\quad
\eqalign{{\cal S}\equiv\s+\half[c,\cdot\,]\ ;\cr
\bar{\cal S}\equiv\sb+\half[\bar c,\cdot\,]\ .}}

In this notation, observe that our field equations \eSTranfs,
\eSbTranfs, \eLambdaI\ and \eLambdaII\ may be compactly rewritten
as
\eqn\eCalEqn{\eqalign{{\cal D}A=F\ ,\cr {\cal S}c=\phi\ ,\cr
\bar{\cal S}\bar c=\bar\phi\ ,\cr
{\cal D}c+{\cal S}A=\psi\ ,\cr
{\cal D}\bar c+\bar{\cal S}A=\bar\psi\ ,\cr
{\cal S}\bar c+\bar{\cal S}c=\lambda\ ,\cr\cr}\qquad
\eqalign{{\cal D}F=0\ ,\cr {\cal S}\phi=0\ ,\cr
\bar{\cal S}\bar\phi=0\ ,\cr\cr\cr\cr\cr}
\qquad\eqalign{{\cal S}F+{\cal D}\psi&=0\ ,\cr
\bar{\cal S}F+{\cal D}\bar\psi&=0\ ,\cr
{\cal D}\phi+{\cal S}\psi&=0\ ,\cr
{\cal D}\bar\phi+\bar{\cal S}\bar\psi&=0\ ,\cr
{\cal S}\lambda+\bar{\cal S}\phi&=0\ ,\cr
\bar{\cal S}\lambda+{\cal S}\bar\phi&=0\ ,\cr
{\cal D}\lambda+{\cal S}\bar\psi&+\bar{\cal S}\psi=0\ .\cr}}
It can be easily recognised that the first column of six
equations represents the curvature equation \eBianchiI a,
while the other two columns describe the Bianchi identity
\eBianchiI b. These equations have an attractive pictorial
representation in Fig.~1. The triplet $(A,c,\bar c)$ is
considered fundamental, out of which all the other fields
are constructed from. Taking one of the sextet fields
$(F,\psi,\bar\psi,\phi,\bar\phi,\lambda)$, it may be expressed
as the sum of its four adjacent fields, each of which is
acted upon by one of the operators $({\cal D},{\cal S},
\bar{\cal S})$. For example, $\psi$ lies next to the fields
$A$ and $c$. Hence, it may be written as $\psi={\cal S}A+
{\cal D}c$. This accounts for the first column.

The other two columns express the fact that all other fundamental
fields outside the triplet and sextet vanish. Take for example,
the position in Fig.~1 with ghost number $+1$ and form degree 2.
It may be written, by the procedure outlined above, as
${\cal S}F+{\cal D}\psi$. But then by \eCalEqn, this field is
identically zero.

It is not clear to us whether or not the simplicity of the
equations in \eCalEqn\ is trying to tell us something else.
For example, while ${\cal D}^2=0$, the equations do not seem to
imply that ${\cal S}^2={\cal S}\bar{\cal S}+\bar{\cal S}{\cal S}
={\cal S}{\cal D}+{\cal D}{\cal S}=\cdots=0$, despite first
appearances. Is it possible to construct from this some sort
of superspace version of (gauge covariant) de Rham cohomology?
Perhaps we should not take \eCalEqn\ too seriously in the first
place, as the dual meanings of ${\cal D}$, ${\cal S}$ and
$\bar{\cal S}$ may be rather misleading.

\newsec{Recovery of standard results}

We have thus so far in this paper, built the solid foundations
of the BRS and anti-BRS symmetry into topological Yang--Mills
theory, and ended with a few speculative remarks. We will now
concentrate, in the rest of this paper, on reproducing the work
of Baulieu and Singer \rBS\ using our new formalism.

Let us define the following auxiliary fields: $b$, an even scalar
field with vanishing ghost number; $\kappa$, an odd one-form with
vanishing ghost number; an odd scalar field $\eta$ with ghost
number one; and its corresponding anti-ghost $\bar\eta$.
Considering the four equations in \eLambdaI\ and \eLambdaII, we
set
\eqn\eAuxI{\eqalign{\s\bar c&=b\ ,\cr \s\bar\psi&=-\kappa\ ,\cr
 \s\lambda&=\eta\ ,\cr \s\bar\phi&=\bar\eta\ ,}
\qquad\eqalign{\sb c&=\lambda-b-[c,\bar c]\ ,\cr
 \sb\psi&=-\D\lambda+\kappa-[c,\bar\psi]-[\bar c,\psi]\ ,\cr
 \sb\lambda&=-\bar\eta-[\bar c,\lambda]-[c,\bar\phi]\ ,\cr
 \sb\phi&=-\eta-[\bar c,\phi]-[c,\lambda]\ .}}
Of course, we could have made a more symmetrical choice of these
transformations, but it does not really matter in our later
considerations.

To ensure the continued nilpotency of $\s$ and $\sb$, we have to
derive the appropriate $\s$ and $\sb$ transformations on the
auxiliary fields. Clearly,
\eqn\eAuxII{\s b=\s\kappa=\s\eta=\s\bar\eta=0\ .}
However, $\sb$ acting on these fields is more complicated
because of our choice of asymmetry in \eAuxI, and we will not
write them down here.

Thus, we have demonstrated how the missing partner of $\bar\eta$
in the BRS formalism naturally appears when we include the
anti-BRS symmetry. We have also managed to reproduce the first
and third row of equations in \eSTranfsI, within our BRS and
anti-BRS formalism. What about the second row of equations,
involving the fields $\bar\chi$ and $B$? Recall that in ref.~\rBS,
the two-form $\bar\chi$ is regarded as the anti-ghost of $\psi$.
But in our analysis of anti-BRS symmetry, the actual anti-ghost
is the one-form $\bar\psi$. Luckily, the reconciliation of this
discrepancy is fairly obvious; these two fields are related by
\eqna\eAntiGh
$$\bar\chi=\d\bar\psi\ ,\eqno\eAntiGh a$$
and similarly for the auxiliary fields:
$$B=\d\kappa\ .\eqno\eAntiGh b$$
(By our choice of gauge-fixing conditions later, $\bar\chi$
and $B$ will be both self-dual two-forms.) While the choice of
these relationships is not unique, it will become apparent later
why we have made the most natural choice. Thus, the second row
of equations in \eSTranfsI\ follows from our analysis as well.

Hence, we have explained the few questions raised earlier,
that is on why the anti-ghost $\bar\chi$ of $\psi$ is not a
one-form; and on the missing partner of $\bar\eta$. This happily
demonstrates the conceptual power and beauty of our combined BRS
and anti-BRS approach.

Our final task is to derive the complete gauge-fixed quantum
action of topological Yang--Mills theory as written down in
ref.~\rBS. To do so, it is useful to first translate our $\s$
and $\sb$ transformations \eSTranfs, \eSbTranfs\ and \eAuxI,
from differential forms into tensor notation. This is an exercise
left to the reader. The total gauge-fixed action consists of
the classical action \eClassicalTYM\ plus an $\s$- and
$\sb$-exact part. It is of the form
\eqn\eGaugeFix{\int_{M}\d^4x\ \tr\,\Bigl[\,F_{\mu\nu}\ast
F^{\mu\nu} + \s\sb\,\{\cdots\}\Bigr]\,\ ,}
for any choice of appropriate gauge-fixing terms within the curly
brackets (with vanishing ghost number). The resulting quantum
action will then be $\s$- and $\sb$-invariant, because of the
nilpotency condition \eNilpotent{}. Recall that $\ast$ is the
duality operator, given by $\ast F_{\mu\nu}={1\over2}\epsilon
_{\mu\nu\rho\sigma}F^{\rho\sigma}$.

With this quantum action $I$, we can define the partition
function
\eqn\ePartFun{Z=\int {\cal D}X \exp(-I/e^2)\ ,}
where ${\cal D}X$ denotes the path integral over the appropriate
fields present in $I$, and $e$ is the coupling constant. Witten
\rTQFT\ has showed how this and suitable correlation functions
of it generate the Donaldson invariants of smooth four-manifolds.
Also recall that because topological field theories are generally
independent of the value of the coupling constant $e$, we can
take the semi-classical limit of very small $e$ \rTQFT. In this
case, only the quadratic terms of $I$ are retained, and any
higher-order terms drop out. This means that we can ignore the
bracket terms in our $\s$ and $\sb$ transformations.

Now recall that the s-exact gauge-fixing part of Baulieu and
Singer's \rBS\ action has the form
\eqn\eGaugeFixI{\s\,\{\bar\chi_{\mu\nu}(F^{\mu\nu}\pm\ast
 F^{\mu\nu})\pm\half\rho\bar\chi_{\mu\nu}
  B^{\mu\nu}+\bar\phi\del_\mu\psi^\mu+\bar c\del_\mu A^\mu
   +\half\sigma\bar c b\}\ ,}
where $\rho$ and $\sigma$ are arbitrary real gauge parameters.
But observe that
\eqnn\eGaugeFixII
$$\eqalignno{\s\sb\,\{\half F_{\mu\nu}F^{\mu\nu}
 &-\half A_\mu A^\mu+\bar\psi_\mu\psi^\mu\} \cr
&=\s\,\{2\del_{[\mu}\bar\psi_{\nu]}F^{\mu\nu}
 +\bar\phi\del_\mu\psi^\mu+\bar c\del_\mu A^\mu
 +\bar\psi_\mu(\del^\mu\lambda-\kappa^\mu-A^\mu)\}\ .
&\eGaugeFixII}$$
After the field redefinition
\eqn\eRedef{\del^\mu\lambda\ \rightarrow\ \del^\mu\lambda
+\kappa^\mu+A^\mu\ ,}
we have
\eqnn\eGaugeFixIII
$$\eqalignno{\s\sb\,\{\half F_{\mu\nu}F^{\mu\nu}
 &-\half A_\mu A^\mu+\bar\psi_\mu\psi^\mu\}\cr
&=\s\,\{2\del_{[\mu}\bar\psi_{\nu]}
 F^{\mu\nu}+\bar\phi\del_\mu\psi^\mu+
 \bar c\del_\mu A^\mu-\lambda\del_\mu\bar\psi^\mu\}\ .
&\eGaugeFixIII}$$

{}From this expression, let us now make the observation that
the $A_\mu$ gauge-fixing condition is $\del_\mu A^\mu=0$;
and that the $\psi_\mu$ gauge-fixing condition is
$\del_\mu\psi^\mu=0$. The apparent gauge-fixing condition
for $F_{\mu\nu}$ is $\del_\mu F^{\mu\nu}=0$, but it is a bad
choice. This is because it is just the Bianchi identity modulo
a higher-order term, and can be made always true in this context.
Instead, the usual choice of gauge for $F_{\mu\nu}$ is either
the self-duality or anti-self-duality condition imposed on it.
This is set by replacing \eGaugeFixII\ with
\eqnn\eGaugeFixIV
$$\eqalignno{\s\sb\,\{\half F_{\mu\nu}P_\pm F^{\mu\nu}
 &-\half A_\mu A^\mu+\bar\psi_\mu\psi^\mu\}\cr
&=\s\,\{2\del_{[\mu}\bar\psi_{\nu]}P_\pm F^{\mu\nu}
 +\bar\phi\del_\mu\psi^\mu+\bar c\del_\mu A^\mu
 -\lambda\del_\mu\bar\psi^\mu\}\ ,&\eGaugeFixIV}$$
where $P_\pm$ is the (anti-) self-dual projection operator
given by $P_\pm\equiv{1\over2}(1\pm\ast)$. Enforcing the
gauge $P_\pm F_{\mu\nu}=0$ is the ``anti-ghost'' field
$\del_{[\mu}\bar\psi_{\nu]}$, which we may conveniently
rename $\bar\chi_{\mu\nu}$. It is an anti-symmetric and
(anti-) self-dual rank-two tensor. At the same time, we set
$B_{\mu\nu}=\del_{[\mu}\kappa_{\nu]}$.

Observe now that \eGaugeFixIV\ is very nearly the same as
\eGaugeFixI, but for the choice of gauge $\rho=\sigma=0$. There
is, however, one extra term
\eqn\eExtraTerm{\s\,\{-\lambda\del_\mu\bar\psi^\mu\}=\lambda
\del_\mu\kappa^\mu-\eta\del_\mu\bar\psi^\mu\ ,}
that does not occur in the latter equation. Fortunately, the
path integrals over the fields $\eta$ and $\lambda$ yield
delta functions which enforce the conditions that
\eqn\eExtraTermI{\del_\mu\bar\psi^\mu=0\ ,\qquad \del_\mu
\kappa^\mu=0\ .}
Note that these do not affect the definitions of $\bar\chi_{\mu
\nu}$ and $B_{\mu\nu}$. Hence, we arrive at the gauge-fixed
action of Baulieu and Singer \rBS, up to negligible higher-order
terms.

The astute reader would notice that in making the field
redefinitions \eAntiGh{}, we are changing the measure of the
path integral by Jacobian-like terms. Symbolically, these
changes are
\eqn\eJacobian{\eqalign{{\cal D}\bar\psi&={\cal D}\bar\chi
 \left[\det{{\cal D}\bar\psi\over{\cal D}\bar\chi}\right]^{-1},
 \cr{\cal D}\kappa&={\cal D} B\left[\det
{{\cal D}\kappa\over{\cal D}B}\right]\ ,}}
where there is an extra inverse in the first Jacobian because
$\bar\psi_\mu$ and $\bar\chi_{\mu\nu}$ are anti-commuting fields.
{}From the relations in \eAntiGh{}, observe that the two Jacobians
cancel each other. Hence the measure of the path integral remains
the same even after our change of variables.

\newsec{Concluding remarks}

By standard arguments, it can be shown that from the gauge
symmetry of a classical theory, one can always build up the
corresponding BRS and anti-BRS symmetry of the quantum theory
counterpart. Thus, in this paper, we have studied the
quantisation of topological Yang--Mills theory, from the point
of view of both the BRS and anti-BRS symmetry. This procedure
explains the various peculiarities that occur in previous
treatments, which consider only the BRS symmetry. In particular,
we have resolved the issue of why the anti-ghost of the ghost
vector field $\psi$ is not a vector field.

Another conceptual advantage of our approach is that it gives a
beautiful geometrical interpretation of the ghost and anti-ghost
fields occurring in our quantisation process. They turn out
to be the connection and curvature components in the two
unphysical, anti-commuting directions of superspace. In
particular, ordinary Yang--Mills theory is recovered by
imposing the condition that these curvature components vanish.
We have also uncovered a certain triality between these two
unphysical directions of superspace, and the physical direction
itself. This triality seems to be described by the Lie group SU(3).

Finally, we showed how to recover the standard gauge-fixed
topological Yang--Mills action from our formalism. We do not
claim that our treatment simplifies this gauge-fixing procedure,
as it clearly does not! Instead, our aim in this paper has been
to demonstrate how to incorporate the anti-BRS symmetry into
topological Yang--Mills theory, and highlight the power of this
method in revealing the elegant geometry and symmetries of the
theory.

One is entitled to ask whether the presence of this extra
anti-BRS symmetry could be used to modify Witten's topological
Yang--Mills theory in any way. Indeed, it is easy to write
down the generalised descent equation \rBS\ which includes
the $\sb$ operator:
\eqn\eDescent{(\d+\s+\sb)(F+\psi+\bar\psi+\phi+\bar\phi
+\lambda)^n=0\ ,}
where $n$ is an integer greater than or equal to 2. By expanding
this equation out in form degree and ghost number, it is in
principle possible to construct new observables of topological
Yang--Mills theory with the appropriate ghost number (see for
example, sec.~5.2.7 of ref.~\rBBRT).

Let us point out another possible extension of this work. As
we have seen, the BRS and anti-BRS symmetry can be interpreted
in terms of connections and curvatures in superspace.
It would be very pleasing if we could write the gauge-fixed
action, \eGaugeFix\ with \eGaugeFixIV, entirely and covariantly
in terms of superfields, like ${\cal A}$ and ${\cal F}$ introduced
earlier. If done, this would no doubt simplify the action and make
any symmetries of the theory more manifest. We do not attempt it
here however, because no entirely satisfactory superfield
formulation yet exists even for ordinary Yang--Mills theory.
Recent and interesting attempts involving the ordinary Yang--Mills
theory may be found in refs.~\ref\rHull{C.M.~Hull, B.~Spence and
J.L.~V\'azquez-Bello, Nucl.~Phys. B348 (1991) 108}\ and
\ref\rGates{S.J.~Gates, {\it in} Strings '90, ed. M.~Duff {\it
et.~al.} (World Scientific, Singapore, 1991)}. Attempts to
describe topological Yang--Mills theory in terms of superfields
have been made in refs.~\ref\rHorne{J.H.~Horne, Nucl.~Phys. B318
(1989) 22}\ and \ref\rAB{C.~Arag\~ ao de Carvalho and L.~Baulieu,
Phys.~Lett. B275 (1992) 323}. To find a complete superfield
formalism would surely be an interesting exercise for the
motivated reader.

Finally, we should mention that anti-BRS symmetry in topological
Yang--Mills theory has also been considered very recently in
ref.~\ref\rPark{H.-J.~Lee and J.-S.~Park, Phys.~Lett. B277
(1992) 119}. However, the structure of their symmetry is very
different from ours. In particular, they have constructed their
BRS and anti-BRS symmetry so that it reproduced the equations
\eSTranfs, where the $\s$ operator consists of their BRS and
anti-BRS operators added together. By contrast, our approach
assumes that \eSTranfs\ alone characterises the BRS part, and
there exists a separate anti-BRS part as in \eSbTranfs. Indeed,
we have followed what is usually done in ordinary Yang--Mills
theory \rBaulieuPR. The reader is invited to compare and contrast
the two approaches.

\bigbreak\bigskip\bigskip\centerline{{\bf Acknowledgement}}
\nobreak
E.T.~wishes to thank The Loke Cheng-Kim Foundation, of
Singapore, for its continued financial support.

\listrefs
\bye